\documentclass[aps,a4paper,superscriptaddress,preprintnumbers,showpacs,
amsmath,amssymb]{revtex4}
\usepackage{graphicx}
\usepackage{dcolumn}
\usepackage{color}
\usepackage{latexsym,amsfonts}
\usepackage{textcomp}
\usepackage{bm}

\usepackage{ulem}

\begin{document}

\title{Exact Solution for Chameleon Field,\\ Self--Coupled Through the
  Ratra--Peebles Potential with $n = 1$\\ and Confined Between Two
  Parallel Plates }

\author{A. N. Ivanov}\email{ivanov@kph.tuwien.ac.at}

\affiliation{Atominstitut, Technische Universit\"at Wien, Stadionallee
  2, A-1020 Wien, Austria}
\author{G.~Cronenberg}\email{cronenberg@ati.ac.at}
\affiliation{Atominstitut, Technische Universit\"at Wien, Stadionallee
  2, A-1020 Wien, Austria}
\author{R.~H\"ollwieser}\email{roman.hoellwieser@gmail.com}\affiliation{Atominstitut,
  Technische Universit\"at Wien, Stadionallee 2, A-1020 Wien,
  Austria}\affiliation{Department of Physics, New Mexico State
  University, Las Cruces, New Mexico 88003, USA}
\author{T.~Jenke}\email{jenke@ill.fr} \affiliation{Institut
  Laue-Langevin, 71 avenue des Martyrs, 38000 Grenoble,
  France}\affiliation{Atominstitut, Technische Universit\"at Wien,
  Stadionallee 2, A-1020 Wien, Austria}
\author{M. Pitschmann}\email{pitschmann@kph.tuwien.ac.at}
\affiliation{Atominstitut, Technische Universit\"at Wien, Stadionallee
  2, A-1020 Wien, Austria}
\author{M. Wellenzohn}\email{max.wellenzohn@gmail.com}
\affiliation{Atominstitut, Technische Universit\"at Wien, Stadionallee
  2, A-1020 Wien, Austria} \affiliation{FH Campus Wien, University of
  Applied Sciences, Favoritenstra\ss e 226, 1100 Wien, Austria}
\author{H. Abele}\email{abele@ati.ac.at} \affiliation{Atominstitut,
  Technische Universit\"at Wien, Stadionallee 2, A-1020 Wien, Austria}
\date{\today}

\begin{abstract}
We calculate the chameleon field profile, confined between two
parallel plates, filled with air at pressure $P = 10^{-4}\,{\rm mbar}$
and room temperature and separated by the distance $L$, in the
chameleon field theory with Ratra--Peebles self--interaction potential
with index $n = 1$.  We give the exact analytical solution in terms of
Jacobian elliptic functions, depending on the mass density of the
ambient matter. The obtained analytical solution can be used in
qBounce experiments, measuring transition frequencies between quantum
gravitational states of ultracold neutrons and also for the
calculation of the chameleon field induced Casimir force for the
CANNEX experiment. We show that the chameleon--matter interactions
with coupling constants $\beta \le 10^4$ can be probed by qBounce
experiments with sensitivities $\Delta E \le 10^{-18}\,{\rm eV}$. At
$L = 30.1\,{\rm \mu m}$ we reproduce the result $\beta < 5.8\times
10^8$, obtained by Jenke {\it et al.}  Phys. Rev. Lett. {\bf 112},
151105 (2014)) at sensitivity $\Delta E \sim 10^{-14}\,{\rm eV}$. In
the vicinity of one of the plates our solution coincides with the
solution, obtained by Brax and Pignol (Phys. Rev. Lett. {\bf 107},
111301 (2011)) (see also Ivanov {\it et al.} Phys. Rev. D {\bf 87},
105013 (2013)) above a plate at zero density of the ambient matter.
\end{abstract}

\maketitle

\section{Introduction}
\label{sec:introduction}

The chameleon field, changing its mass in dependence of the density of
its environment \cite{Chameleon1,Chameleon2}, has been invented to
avoid the problem of the equivalence principle violation
\cite{Will1993}. Nowadays it is accepted that the chameleon field,
identified with quintessence
\cite{Zlatev1999,Tsujikawa2013,Ivanov2016a}, i.e. a canonical scalar
field, can be useful for the explanation of the late--time
acceleration of the Universe expansion
\cite{Perlmutter1997,Riess1998,Perlmutter1999,Goobar2000} (see, for
example, \cite{Brax2004}). In addition, the chameleon field may shed
light on dark energy dynamics
\cite{Peebles2003}--\cite{Pignol2015}. In terrestrial laboratories
\cite{Abele2010}--\cite{Li2016} chameleon--matter interactions have
been investigated in terms of ultracold neutrons, coupled to the
chameleon field above a mirror \cite{Brax2011} and between two mirrors
\cite{Ivanov2013}, as well as cold neutrons by using neutron
interferometers \cite{Brax2013,Lemmel2015,Li2016} and via atom
interferometry \cite{Burrage2014,Burrage2015} with ${^{133}}{\rm Cs}$
atoms as test particles, \cite{Hamilton2015,Hamilton2016} in the
chameleon field theory with the Ratra--Peebles self--interaction
potential.  In these experiments the contribution of the chameleon
field has been extracted in terms of the chameleon--matter coupling
constant $\beta$ with the upper bounds $\beta < 5.8 \times 10^8$
(qBounce experiment for Ratra--Peebles potential with index $1\le n
\le 10$) \cite{Jenke2014}, $\beta < 1.9\times 10^7$ (neutron
interferometry for the Ratra--Peebles potential with index $n = 1$)
\cite{Lemmel2015} and $\beta < 4.5 \times 10^4$ (atom interferometry
for Ratra--Peebles potential with index $n = 1$)
\cite{Hamilton2015,Hamilton2016}. Unfortunately, there is still no
plausible explanation why the chameleon field couples to neutrons as
test particles with strength $\beta < 10^7$, whereas to ${^{133}}{\rm
  Cs}$ test particles with $\beta < 10^4$. The new run of qBounce
experiments with ultracold neutrons, using Gravity Resonance
Spectroscopy and invented for measurements of the transitions
frequencies between quantum gravitational states of ultracold neutrons
bouncing above a mirror \cite{Gunther2016}, should give new upper
bounds on the chameleon--neutron couplings.

In this paper we give the analytical solution for the chameleon field,
confined between two parallel plates, in the chameleon field theory
with Ratra-- Peebles self--interaction potential with index $n =
1$. Such a potential is very popular in atom interferometry
\cite{Burrage2014,Burrage2015,Hamilton2015,Hamilton2016}. In section
\ref{sec:equation} we write the equation of motion for the chameleon
field and define the effective potential of the chameleon--matter and
chameleon self--interactions. In section \ref{sec:solution} we solve
the equation of motion for the chameleon field profile in terms of
Jacobian elliptic functions. In section \ref{sec:phi0} we use the
boundary conditions for the chameleon field and derive the equation
for the definition of the parameter $\phi_0$, which is the maximum
value of the chameleon field between two parallel plates.  We make
numerical calculations of the parameters of the chameleon field
profile. In section \ref{sec:transition} we calculate the
contributions of the chameleon field profile, derived in section
\ref{sec:solution}, to the transition frequencies of quantum
gravitational states of ultracold neutrons. In section
\ref{sec:region} we discuss the allowed region of the
chameleon--matter coupling constant$\beta$, i.e. $0 \le \beta <
\beta_{\rm max}$, which can be measured experimentally for the
chameleon field coupled to neutron, confined between two parallel
plates. We show that the maximum value $\beta_{\rm max}$ depends on
the density of the ambient matter $\rho$ and the distance $L$ between
plates. Such a dependence appears because of the not
positive--definite of the first integral of the equation of motion of
the chameleon field between two parallel plates. Indeed, the squared
first derivative of the chameleon field is proportional to the
difference $V_{\rm eff}(\phi) - V_{\rm eff}(\phi_0)$ of the effective
potentials, which is not positive--definite for an arbitrary product
$\beta \rho$. Together with the boundary condition Eq.(\ref{eq:21}),
which depends on the distance between two parallel plates $L$, this
imposes the constraint on $\beta_{\rm max}$ in dependence of the
density of the ambient matter $\rho$ and the distance $L$. Of course,
the obtained constraints on $\beta_{\rm max}$ are valid only for
$1D$--profiles of the chameleon field between two parallel plates and
violated for $2D$-- and $3D$--profiles. We show that the
chameleon--matter coupling constant $\beta \le 10^4$ can be extracted
from the experimental data on the transition frequencies of quantum
gravitational states of ultracold neutrons for experimental
sensitivities $\Delta E \le 10^{-18}\,{\rm eV}$. We would like to
emphasize that currently such a sensitivity is not reachable at the
qBounce experiments. Presently one may relay on the sensitivity
$\Delta E \le 5\times 10^{-15}\,{\rm eV}$ \cite{Gunther2016}. However,
according to Abele {\it et al.} \cite{Abele2010}, one may expect soon
a substantial improvement of the sensitivity up to $\Delta E \le
10^{-17}\,{\rm eV}$ (Phase I Ramsey Experiment) and to $\Delta E \le
10^{-21}\,{\rm eV}$ (Phase II Ramsey Experiment) \cite{Jenke2014}. In
section \ref{sec:conclusion} we discuss the obtained results.

\section{Equation of motion for chameleon field, confined between 
two plates}
\label{sec:equation}

In this section we search for the solution of the chameleon equation
of motion for the chameleon field, confined between two plates
parallel the $(x,y)$ plane, separated by a length $L$ and localized
at $z = 0$ and $z = L$, respectively. Following \cite{Ivanov2013}
the corresponding equation of motion for the chameleon field $\phi(z)$
is given by
\begin{align}\label{eq:1}
\frac{d^2 \phi}{d z^2} = \frac{\partial V_{\rm
    eff}(\phi)}{\partial \phi},
\end{align}
where $V_{\rm eff}(\phi)$ is the effective potential defined by 
\begin{align}\label{eq:2}
V_{\rm eff}(\phi) = \frac{\Lambda^5}{\phi} + \beta\,\frac{\rho}{M_{\rm
    Pl}}\,\phi.
\end{align}
Here $\Lambda^5/\phi$ is the Ratra--Peebles potential of the
self--interaction of the chameleon field with index $n = 1$, whereas
the term proportional to the matter density $\rho$ is the potential of
the chameleon--matter interaction. Then, $\Lambda = \sqrt[4]{3
  M^2_{\rm Pl}{\rm H}^2_0\,\Omega_{\Lambda}} = 2.24(2)\times
10^{-3}\,{\rm eV}$ is the dark energy scale parameter, $M_{\rm Pl} =
1/\sqrt{8\pi G_N} = 2.435\times 10^{27}\,{\rm eV}$ and $G_N$ are the
reduced Planck mass and the Newtonian gravitational constant,
respectively, and $\Omega_{\Lambda} \simeq 0.685$ is the relative dark
energy density in the Universe \cite{PDG2014}.

\section{Solution to equation of motion for chameleon field, 
confined between two plates}
\label{sec:solution}

The solution of Eq.~(\ref{eq:1}) we obtain from the first integral of
Eq.~(\ref{eq:1}) equal to
\begin{align}\label{eq:3}
\frac{1}{2}\,\Big(\frac{d\phi}{d z}\Big)^2 = V_{\rm
  eff}(\phi) - V_{\rm eff}(\phi_0),
\end{align}
where $V_{\rm eff}(\phi_0)$ is the integration constant, given by
\begin{align}\label{eq:4}
V_{\rm eff}(\phi_0) = \frac{\Lambda^5}{\phi_0} +
\beta\,\frac{\rho}{M_{\rm Pl}}\,\phi_0,
\end{align}
and $\phi_0$ is the value of the chameleon field at $z = L/2$,
i.e. $\phi(L/2) = \phi_0$ and $d\phi(z)/dz|_{z = L/2} = 0$
\cite{Ivanov2013}. The solution to Eq.~(\ref{eq:3}) we obtain in the
form of the integral \cite{Ivanov2013}
\begin{align}\label{eq:5}
\frac{\phi_0^{3/2}}{\Lambda^{5/2}}\int^1_x \frac{\sqrt{x'}\,dx'}{\sqrt{(1
    - x')(1 - k^2 x')}} = \sqrt{2}\,\Big(\frac{L}{2} - z\Big),
\end{align}
where $x = \phi/\phi_0$ and $k^2 = \phi^2_0/\phi^2_{\rm v}$ with
$\phi_{\rm v}$ equal to $\phi_{\rm v} = \sqrt{\Lambda^5M_{\rm
    Pl}/\beta \rho}$ \cite{Ivanov2013}. Here $\phi_{\rm v}$ is a
minimum of the chameleon field, which it can reach in the unrestricted
space interval. Making a change of variables $x' = \sin^2\!\varphi$ we
arrive at the expression
\begin{align}\label{eq:6}
\int^{\pi/2}_{\arcsin\sqrt{x}} \frac{\sin^2\!\varphi\,d\varphi}{\sqrt{1
    - k^2\sin^2\!\varphi}} = \frac{2}{L_S}\Big(\frac{L}{2} - z\Big),
\end{align}
where $L_S = \sqrt{2}\,\phi^{3/2}_0/\Lambda^{5/2}$ is the slope
parameter. The right--hand--side (r.h.s.) of Eq.~(\ref{eq:6}) we give
in the form of the derivative with respect to the parameter
$k^2$. Replacing then $k^2 \to p^2$ we get
\begin{align}\label{eq:7}
\frac{\partial}{\partial p^2}\int^{\pi/2}_{\arcsin\sqrt{x}}\sqrt{1 - p^2 \sin^2\!\varphi}\,d\varphi = - \frac{1}{L_S}\Big(\frac{L}{2} - z\Big).
\end{align}
Integrating Eq.~(\ref{eq:7}) over $p^2$ in the limits $0 \le p^2 \le
k^2$ we arrive at the expression
\begin{align}\label{eq:8}
\int^{\pi/2}_{\arcsin\sqrt{x}}\sqrt{1 - k^2 \sin^2\!\varphi}\,d\varphi -
\arccos\sqrt{x} = - \frac{k^2}{L_S}\Big(\frac{L}{2} - z\Big).
\end{align}
The integral over $\varphi$ can be given in terms of the elliptic
integral of the second kind \cite{HMF1972}
\begin{align}\label{eq:9}
\int^{\pi/2}_{\arcsin\sqrt{x}}\sqrt{1 - k^2 \sin^2\!\varphi}\,d\varphi =
E\Big(\frac{\pi}{2},k\Big) - E(\arcsin\sqrt{x},k).
\end{align}
This allows to rewrite  Eq.~(\ref{eq:8}) as follows
\begin{align}\label{eq:10}
E(\arcsin\sqrt{x},k) + \arccos\sqrt{x} = E\Big(\frac{\pi}{2},k\Big) +
\frac{k^2}{L_S}\Big(\frac{L}{2} - z\Big).
\end{align}
Then, we use the following properties of the elliptic integral of the second kind \cite{HMF1972a}
\begin{align}\label{eq:11}
E(u,k) = u - k^2\int^u_0 {\rm sn}^2(t,k)\,dt = u - k^2\,{\rm Sn}(u,k),
\end{align}
where ${\rm sn}(t,k)$ is the Jacobian elliptic function
\cite{HMF1972}. Plugging Eq.~(\ref{eq:11}) into Eq.~(\ref{eq:10}) we get
\begin{align}\label{eq:12}
{\rm Sn}\Big(\arcsin\sqrt{\frac{\phi(z)}{\phi_0}},k\Big) = {\rm
  Sn}\Big(\frac{\pi}{2},k\Big) - \frac{1}{L_S}\Big(\frac{L}{2} -
z\Big),
\end{align}
where we have set $x = \phi(z)/\phi_0$ (see Eq.~(\ref{eq:6})).  The
chameleon field $\phi(z)$ is given by
\begin{align}\label{eq:13}
\phi(z) = \phi_0 \sin^2\Big({\rm Sn}^{-1}\Big\{\Big[{\rm
    Sn}\Big(\frac{\pi}{2},k\Big) - \frac{1}{L_S}\Big(\frac{L}{2} -
  z\Big)\Big],k\Big\}\Big),
\end{align}
where ${\rm Sn}^{-1}(x, k)$ is the inverse function of the Jacobian
elliptic function ${\rm Sn}(u,k) = x$. Since at $z = L/2$ we get ${\rm
  Sn}^{-1}\{{\rm Sn}(\pi/2,k),k\} = \pi/2$, the solution
Eq.~(\ref{eq:13}) gives $\phi(L/2) = \phi_0$. 

Practically, the solution Eq.~(\ref{eq:13}) is valid in the space
interval $0 \le z \le L/2$. For the definition of the solution valid
in the space interval $0 \le z \le L$ we follow \cite{Ivanov2013} and
get
\begin{align}\label{eq:14}
\phi(z) = \phi_0 \sin^2\Big({\rm Sn}^{-1}\Big\{\Big[{\rm
    Sn}\Big(\frac{\pi}{2},k\Big) - \frac{L}{2 L_S}\Big|1
  -\frac{2z}{L}\Big|\Big],k\Big\}\Big),\quad \quad 0 \le z \le L.
\end{align}
Let us show that the solution Eq.~(\ref{eq:14}) is continuous in the
vicinity of $z = L/2$. For this aim we define the chameleon field as
follows $\phi(z) = \phi_0 - \delta \phi(z)$, where $\delta \phi(z)$ is
a small deviation of the chameleon field $\phi(z)$ from $\phi_0$ in a
small vicinity of $z = L/2$ such as $\phi_0 \gg \delta \phi(z)$. Using
then the expansion
\begin{align}\label{eq:15}
{\rm Sn}\Big(\arcsin\sqrt{\frac{\phi(z)}{\phi_0}},k\Big) = {\rm
  Sn}\Big(\arcsin\sqrt{1 - \frac{\delta \phi(z)}{\phi_0}},k\Big) =
{\rm Sn}\Big(\frac{\pi}{2},k\Big) - {\rm
  sn}^2\Big(\frac{\pi}{2},k\Big)\sqrt{\frac{\delta \phi(z)}{\phi_0}}
\end{align}
and Eq.~(\ref{eq:12}) we get
\begin{align}\label{eq:16}
\delta \phi(z) = \frac{\phi_0}{\bar{L}^2_S}\Big(\frac{L}{2} -
z\Big)^2,
\end{align}
where $\bar{L}_S = L_S {\rm sn}^2(\frac{\pi}{2},k)$. Thus, in the
vicinity of $z = L/2$ the chameleon field profile is defined by the
function
\begin{align}\label{eq:17}\phi(z) = \phi_0 \Big(1 - \frac{L^2}{4\bar{L}^2_S}\Big(1 -
\frac{2z}{L}\Big)^2\Big).
\end{align}
This confirms a continuity of the chameleon field profile
Eq.(\ref{eq:14}) in the vicinity of $z = L/2$.  Now we may proceed to
the definition of the parameter $\phi_0$. For this aim we have to use
the boundary conditions.

\section{Boundary conditions and definition of $\phi_0$}
\label{sec:phi0}

The main parameter of the solution Eq.~(\ref{eq:14}) is $\phi_0$, which
defines also $k = \phi_0/\phi_{\rm v}$.  Following \cite{Ivanov2013}
we determine it from boundary conditions:
\begin{align}\label{eq:18}
&\phi(z)\Big|_{z = 0-} = \phi(z)\Big|_{z = 0+} = \phi_b \quad,\quad
\phi(z)\Big|_{z = L -} = \phi(z)\Big|_{z = L+} =
\phi_b,\nonumber\\ &\Big(\frac{d\phi}{dz}\Big)^2\Big|_{z = 0-} =
\Big(\frac{d\phi}{dz}\Big)^2\Big|_{z = 0+} \quad,\quad
\Big(\frac{d\phi}{dz}\Big)^2\Big|_{z = L-} =
\Big(\frac{d\phi}{dz}\Big)^2\Big|_{z = L+}.
\end{align}
Skipping intermediate calculations (see \cite{Ivanov2013}) we arrive at the relation
\begin{align}\label{eq:19}
\phi_b = \frac{\displaystyle 2 \phi_m - \frac{\rho}{\rho_m}\,\phi_0 -
  \frac{\phi^2_m}{\phi_0}}{\displaystyle 1 - \frac{\rho}{\rho_m}} =
\frac{\displaystyle 2 \phi_m - \frac{\phi^2_m}{\phi^2_{\rm v}}\,\phi_0
  - \frac{\phi^2_m}{\phi_0}}{\displaystyle 1 -
  \frac{\phi^2_m}{\phi^2_{\rm v}}},
\end{align}
where $\rho_m$ is the mass density of the plates and $\phi_m$ is the
minimum of the chameleon field inside the plates $\phi_m =
\sqrt{\Lambda^5 M_{\rm Pl}/\beta\rho_m}$. At $\rho \to 0$ or
$\phi_{\rm v} \to \infty$ Eq.~(\ref{eq:19}) reduces to Eq.~(39) of
Ref.~\cite{Ivanov2013}. One more equation we obtain by using
Eq.~(\ref{eq:13}) at $z = 0$ (or Eq.~(\ref{eq:14}) at $z = L$), which
can be rewritten as follows
\begin{align}\label{eq:20}
{\rm Sn}\Big(\arcsin\sqrt{\frac{\phi_b}{\phi_{\rm v}}\frac{\phi_{\rm
      v}}{\phi_0}},\frac{\phi_0}{\phi_{\rm v}}\Big) = {\rm
  Sn}\Big(\frac{\pi}{2},\frac{\phi_0}{\phi_{\rm v}}\Big) -
\frac{L\Lambda}{2\sqrt{2}}\Big(\frac{\Lambda}{\phi_{\rm
    v}}\Big)^{3/2}\Big(\frac{\phi_{\rm v}}{\phi_0}\Big)^{3/2}.
\end{align}
Plugging Eq.~(\ref{eq:19}) into Eq.~(\ref{eq:20}) we obtain the
equation, which defines the parameter $\phi_0$ through the dark energy
scale $\Lambda$, the spatial scale of the experimental setup $L$ and
matter densities $\rho$ and $\rho_m$ between two parallel plates and
inside the plates, respectively. For numerical calculations we use $L
= 20\,{\rm cm}$, $\rho = 1.188\times 10^{-10}\,{\rm g/cm^3} =
5.123\times 10^8\,{\rm eV^4}$, corresponding the air density and
pressure $P = 10^{-4}\,{\rm mbar}$, $\rho_m = 2.51\,{\rm g/cm^3} =
1.082\times 10^{19}\,{\rm eV^4}$ \cite{Gunther2016}, and $\Lambda =
2.24\times 10^{-3}\,{\rm eV}$ and $\beta = 10^4$. However, in order to
reproduce the result $\beta < 10^7$, obtained in \cite{Gunther2016},
we have to use $L = 1\,{\rm cm}$. The value of the chameleon--matter
coupling constant $\beta = 10^4$ is chosen in agreement with recent
results in the atom interferometry
\cite{Hamilton2015,Hamilton2016}. Since for $\beta = 10^4$, $\Lambda =
2.24\times 10^{-3}\,{\rm eV}$ and $\rho_m = 1.082\times 10^{19}\,{\rm
  eV^4}$ we get $\phi_m = \sqrt{\Lambda^5 M_{\rm Pl}/\beta\rho_m} =
3.56\times 10^{-5}\,{\rm eV}$, for the accepted parameters of the
experimental setup \cite{Gunther2016} we may set $\phi_b = 0$. As a
result, we arrive at the equation for the parameter $k$, given by
\begin{align}\label{eq:21}
k^{3/2}{\rm Sn}\Big(\frac{\pi}{2},k\Big) =
\frac{L\Lambda}{2\sqrt{2}}\Big(\frac{\Lambda}{\phi_{\rm
    v}}\Big)^{3/2}.
\end{align}
For $\beta = 10^4$, $\Lambda = 2.24\times 10^{-3}\,{\rm eV}$ and $\rho
= 5.123\times 10^8\,{\rm eV^4}$ we obtain $\phi_{\rm v} =
\sqrt{\Lambda^5 M_{\rm Pl}/\beta\rho} = 5.177\,{\rm eV}$, and for $L =
20\,{\rm cm}$ the numerical solution of Eq.~(\ref{eq:21}) gives $k =
0.070$ and $\phi_0 = 0.360\,{\rm eV}$.

\begin{figure}
\centering \includegraphics[height=0.18\textheight]{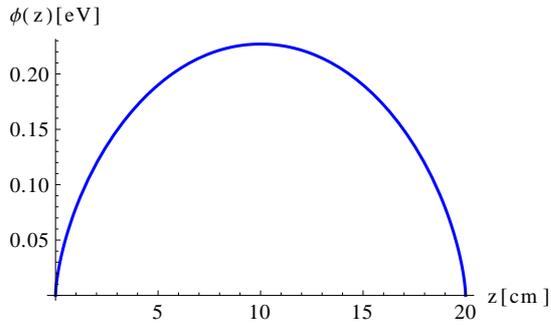}
\caption{The profile of the chameleon field, defined by
  Eq.~(\ref{eq:14}) in the spatial region $0 \le z \le L = 20\,{\rm
    cm}$ between two parallel plates and calculated for $\beta =
  10^4$, air density $\rho = 1.188\times 10^{-10}\,{\rm g/cm^3}$ at
  pressure $P = 10^{-4}\,{\rm mbar}$ and room temperature.}
\label{fig:phi}
\end{figure}

In Fig.\,\ref{fig:phi} we give the profile of the chameleon field,
defined by Eq.~(\ref{eq:14}) in the spatial region $0 \le z \le L =
20\,{\rm cm}$ and calculated for $\beta = 10^4$, air density $\rho =
1.188\times 10^{-10}\,{\rm g/cm^3}$ at pressure $P = 10^{-4}\,{\rm
  mbar}$ and room temperature.

\section{Transition frequencies between quantum gravitational states
 of ultracold neutrons}
\label{sec:transition}

Because of a very large value $L = 20\,{\rm cm}$ the wave functions of
the quantum gravitational states of ultracold neutrons we may take in
the form, corresponding to ultracold neutrons bouncing in the
gravitational field of the Earth above a mirror \cite{Gibbs1975}
\begin{align}\label{eq:22}
\hspace{-0.3in}\psi_{k'}(z) = \frac{1}{\sqrt{\ell_0}}\frac{\displaystyle
  {\rm Ai}(\xi_{k'} + \xi)}{\displaystyle
  \sqrt{\int^{\infty}_0 \Big|{\rm Ai}(\xi_{k'} + \xi)\Big|^2 d\xi}},
\end{align}
where $\xi = z/\ell_0$, ${\rm Ai}(x)$ is the Airy function, $\ell_0 =
(2m^2g)^{-1/3} = 5.87\,{\rm \mu m}$, $\xi_{k'}$ are the roots of the
equation ${\rm Ai}(\xi_{k'}) = 0$, caused by the boundary condition
$\psi_{k'}(0) = 0$. The roots $\xi_{k'}$ define the energy spectrum of
quantum gravitational states $E_{k'} = - m g \ell_0 \xi_{k'}$ for $k'
= 1,2,\ldots$, where $m g \ell_0 = 0.602\,{\rm peV}$.  According
\cite{Gunther2016}, the transitions $|1\rangle \to |3\rangle$ and
$|1\rangle \to |4\rangle$ there have been only observed. The
contributions $\delta \omega_{k' 1}$ of the chameleon field to the
transition frequencies of the transitions $|1\rangle \to |k'\rangle$
are equal to
\begin{align}\label{eq:23}
\hspace{-0.3in}\delta \omega_{k'1} = \beta\,\frac{m}{M_{\rm
    Pl}}\int^{L/\ell_0}_0 d\xi\,\phi(\xi\ell_0)\,\Big(|\psi_{k'}(\xi)|^2
- |\psi_1(\xi)|^2\Big)
\end{align}
for $k' = 3,4$ \cite{Gunther2016}, where $\xi = z/\ell_0$ and the
chameleon field profile $\phi(\xi\ell_0)$ is given by
Eq.~(\ref{eq:14}). The numerical calculations, carried out with the
chameleon field profile Eq.~(\ref{eq:14}), give the following results:
$\delta \omega_{31} = 3.72\times 10^{-18}\,{\rm eV}$ and $\delta
\omega_{41} = 4.98\times 10^{-18}\,{\rm eV}$. Thus, the contribution
of the chameleon field to the transition frequencies of quantum
gravitational states of ultracold neutrons, bouncing above a mirror in
the gravitational field of the Earth, with the chameleon--matter
coupling constant $\beta = 10^4$ are at the level of sensitivities
$\Delta E \sim 10^{-18}\,{\rm eV}$.

For the comparison we propose the following approximation. One may
assert that because of the wave functions of the gravitational states
of ultracold neutrons the contributions to the transition frequencies
$\delta \omega_{31}$ and $\delta \omega_{41}$ come from the spatial
region or order of a few micrometers. This means that we may use an
approximation for the chameleon field profile in the vicinity of the
lower plate. Assuming that $\phi_0 \gg \phi(z)$ we transform
Eq.~(\ref{eq:14}) into the form
\begin{eqnarray}\label{eq:24}
\phi(z) = \Lambda \Big(\frac{3}{\sqrt{2}}\,\Lambda
\ell_0\Big)^{2/3}\,\Big(\frac{z}{\ell_0}\Big)^{2/3} = 6.078\times
10^{-4}\,\Big(\frac{z}{\ell_0}\Big)^{2/3}\,{\rm eV},
\end{eqnarray}
which agrees well with the results, obtained in
\cite{Brax2011,Ivanov2013} for the chameleon field profile above a
mirror in the space with density $\rho = 0$. Plugging
Eq.~(\ref{eq:24}) into Eq.~(\ref{eq:23}) and integrating over the
region $0 \le \xi < \infty$ we obtain $\delta \omega_{31} = 2.34\times
10^{-18}\,{\rm eV}$ and $\delta \omega_{41} = 3.14\times
10^{-18}\,{\rm eV}$ at $\beta = 10^4$.

\section{Allowed region of the chameleon--matter coupling constant 
$0 \le \beta < \beta_{\rm max}$ in dependence of the parameters of the
  experimental setup}
\label{sec:region}

According to Eq.(\ref{eq:5}), the right--hand--side (r.h.s.) $V_{\rm
  eff}(\phi) - V_{\rm eff}(\phi_0)$ of Eq.(\ref{eq:3}) is not
positive--definite. Since the chameleon field is a real scalar field,
the integrand of Eq.(\ref{eq:5}) should be a real function. This
implies that the parameter $k^2 = \phi^2_0/\phi^2_v$ should be
restricted from above by unity, i.e. $k^2 < 1$ or $\phi^2_0 <
\phi^2_{\rm v}$ (see Eqs.(\ref{eq:6}) - (\ref{eq:9})). Together with
the boundary condition Eq.(\ref{eq:21}) this gives the constraint on
the allowed region of the chameleon--matter coupling constant $\beta$
\begin{eqnarray}\label{eq:25}
0 \le \beta < \beta_{\rm max} = \frac{\Lambda^5 M_{\rm Pl}}{\rho
  \phi^2_0}.
\end{eqnarray}
For the air density $\rho = 1.188\times 10^{-10}\,{\rm g/cm^3}$,
corresponding to air density at pressure $P = 10^{-4}\,{\rm mbar}$ and
room temperature, and $\phi_0 = 0.277\,{\rm eV}$, caused by the
solution of the boundary condition Eq.(\ref{eq:21}), we get
$\beta_{\rm max} = 3.5\times 10^6$. For the Ratra--Peebles potential
with index $n$ the allowed region of the chameleon--matter coupling
constant is restricted by
\begin{eqnarray}\label{eq:26}
0 \le \beta < \beta_{\rm max} = \frac{\Lambda^{4 + n} M_{\rm Pl}}{\rho
  \phi^{n + 1}_0}.
\end{eqnarray}
We would like to note that $\beta_{\rm max}$ does not define the upper
bound of the chameleon--matter coupling constant, but it defines a
maximal value of the chameleon--matter coupling constant, which can be
measured for given density $\rho$ and the distance $L$, caused by the
experimental setup.

We would like to emphasize that the constraint on the maximal value of
the chameleon--matter coupling constant exists only for the chameleon
field, confined between two parallel infinitely large plates. In this
case the second order differential equation of motion of the chameleon
field possesses the first integral, having the shape of
Eq.(\ref{eq:3}), where the r.h.s. is not positive--definite. For any
2D-- or 3D--profile of the chameleon field, when the equation of
motion
\begin{eqnarray}\label{eq:27}
\Delta \phi = \frac{\partial V_{\rm eff}(\phi)}{\partial \phi},
\end{eqnarray}
where $\Delta$ is the Laplacian in the 2D-- or 3D--dimensional space,
cannot be reduced to the form of Eq.(\ref{eq:3}), the constraint on
the maximal value of the chameleon--matter coupling constant
$\beta_{\rm max}$ does not exist. The latter concerns the results,
obtained in the neutron interferometry by Brax {\it et al.}
\cite{Brax2013} and Lemmel {\it et al.} \cite{Lemmel2015}, where the
2D-- and 3D--profiles of the chameleon field have been used for the
analysis of the phase shift of the neutron wave function,
respectively.

\section{Conclusion}
\label{sec:conclusion}

We have found the exact solution for the chameleon field, confined
between two parallel plates and described by the chameleon field
theory with the Ratra--Peebles potential with index $n = 1$. Such a
potential is very popular for the measurements of the
chameleon--matter coupling constant in atom interferometry
\cite{Burrage2014,Burrage2015,Hamilton2015,Hamilton2016}. We have
applied the obtained chameleon field profile to the analysis of the
transition frequencies of the quantum gravitational states of
ultracold neutrons, bouncing above a mirror in the gravitational field
of the Earth \cite{Gunther2016}. We have shown that in the vicinity of
the plate at $z = 0$ our exact solution Eq.(\ref{eq:14} reduces to the
form $\phi(z) = \Lambda (3\Lambda
\ell_0/\sqrt{2})^{2/3}(z/\ell_0)^{2/3}$, which agrees well with the
chameleon field profile, calculated in \cite{Brax2011,Ivanov2013} for
zero density of ambient matter. Then, we have shown that the
contribution of the chameleon field profile to the transition
frequencies between the states $|1\rangle \to |3\rangle$ and
$|1\rangle \to |4\rangle$, caused by the chameleon--neutron
interaction with the chameleon--matter coupling constant $\beta \le
10^4$ \cite{Hamilton2015,Hamilton2016}, can be observed only at the
level of sensitivities $\Delta E \le 10^{-18}\,{\rm eV}$.


Replacing $L = 20\,{\rm cm}$ by $L = 30.1\,{\rm \mu m}$ the obtained
profile of the chameleon field Eq.(\ref{eq:14}) can be used for the
qBounce experiments with ultracold neutrons, bouncing between two
mirrors \cite{Jenke2014} and described by the wave functions
\cite{Ivanov2013}, and the measurement of the Casimir force, caused by
the chameleon field, in the CANNEX experiments \cite{Sedmik2016}.

We would like to emphasize that the allowed region $0 \le \beta <
\beta_{\rm max}$ for a measurement of the chameleon--matter coupling
constant $\beta$ depends on the ambient matter density $\rho$ and the
distance $L$ between two parallel plates.

For example, for the air density $\rho = 1.188\times 10^{-10}\,{\rm
  g/cm^3}$ at pressure $P = 10^{-4}\,{\rm mbar}$ and room temperature
and the distance between two plates $L = 30,1\,{\rm \mu m}$
\cite{Jenke2014} the constraint $k < 1$ and the boundary condition
Eq.(\ref{eq:21}) define the following allowed region of the
chameleon--matter coupling constant $0 \le \beta < \beta_{\rm max} =
4\times 10^{11}$. This does not contradict the result $\beta < 5.8
\times 10^8$, obtained by Jenke {\it et al.}  \cite{Jenke2014}. For
the chameleon--matter coupling constant $\beta = 5.8\times 10^8$ and
density $\rho = 1.188\times 10^{-10}\,{\rm g/cm^3}$ we get $\phi_0 =
1.022\times 10^{-3}\,{\rm eV}$, $\phi_{\rm v} = 2.150\times
10^{-2}\,{\rm eV}$ and $k = \phi_0/\phi_{\rm v} = 0.048$. This gives
the following contributions of the chameleon field to the transition
frequencies $\delta \omega_{31} = 5.20\times 10^{-14}\,{\rm eV}$ and
$\delta \omega_{41} = 6.67\times 10^{-14}\,{\rm eV}$, which agree well
with the results obtained by Jenke {\it et al.}  \cite{Jenke2014}. For
the calculation of the transition frequencies $\delta \omega_{31} =
5.20\times 10^{-14}\,{\rm eV}$ and $\delta \omega_{41} = 6.67\times
10^{-14}\,{\rm eV}$ we have used the wave functions of the ultracold
neutrons, confined between two parallel plates and defined in
\cite{Ivanov2013}.

For $L = 100\,{\rm \mu m}$ and $\rho = 1.188\times 10^{-10}\,{\rm
  g/cm^3}$ we get the allowed region of the chameleon--matter coupling
constants $0 \le \beta < \beta_{\rm max} = 8.5\times 10^{10}$. Then,
for $\beta = 10^7$ we obtain $\phi_0 = 2.271\times 10^{-3}\,{\rm eV}$,
$\phi_{\rm v} = 0.164\,{\rm eV}$ and $k = \phi_0/\phi_{\rm v} =
0.014$. The contributions of the chameleon field to the transition
frequencies are equal to $\delta \omega_{31} = - 2.69\times
10^{-17}\,{\rm eV}$ and $\delta \omega_{41} = + 5.45\times
10^{-16}\,{\rm eV}$. These results become observable at sensitivity
$\Delta E < 10^{-17}\,{\rm eV}$ \cite{Abele2010}.

For $L = 1\,{\rm cm}$ and $\rho = 1.188\times 10^{-10}\,{\rm g/cm^3}$
we get the allowed region of the chameleon--matter coupling constants
$0 \le \beta < \beta_{\rm max} = 1.6\times 10^8$. Then, for $\beta =
10^7$ we obtain $\phi_0 = 4.825\times 10^{-2}\,{\rm eV}$, $\phi_v =
0.164\,{\rm eV}$ and $k = \phi_0/\phi_{\rm v} = 0.029$. The
contributions of the chameleon field to the transition frequencies
amount to $\delta \omega_{31} = 1.23\times 10^{-15}\,{\rm eV}$ and
$\delta \omega_{41} = 1.65\times 10^{-15}\,{\rm eV}$. These results
can be observed at sensitivity $\Delta E < 5\times 10^{-15}\,{\rm eV}$
\cite{Gunther2016}.

For the density $\rho = 1.642\times 10^{-11}\,{\rm g/cm^3}$ of helium
gas at pressure $P = 10^{-4}\,{\rm mbar}$ and room temperature, which
has been used by Brax {\it et al.}  \cite{Brax2013}, and $L = 1\,{\rm
  cm}$ we get the allowed region of the chameleon--matter coupling
constant equal to $0 \le \beta < \beta_{\rm max} = 1.3\times
10^9$. Then, for $\beta = 10^7$ we obtain $\phi_0 = 4.885\times
10^{-2}\,{\rm eV}$, $\phi_v = 0.441\,{\rm eV}$ and $k =
\phi_0/\phi_{\rm v} = 0.111$. The contributions of the chameleon field
to the transition frequencies amount to $\delta \omega_{31} =
1.23\times 10^{-15}\,{\rm eV}$ and $\delta \omega_{41} = 1.66\times
10^{-15}\,{\rm eV}$. These results can be also observed at sensitivity
$\Delta E < 5\times 10^{-15}\,{\rm eV}$ \cite{Gunther2016}.

Finally we would like to mention the paper by Burrage, Copeland, and
Stevenson \cite{Burrage2016}, where the authors solve the problem of
the chameleon field, confined between two parallel plates.  Unlike our
solution Eq.(\ref{eq:14}), calculated for a non-vanishing matter
density $\rho \neq 0$ between two parallel plates, the solution
Eq.(\ref{eq:5}) is calculated in \cite{Burrage2016} at zero matter
density $\rho = 0$. It is important to emphasize that because of the
constraint $k^2 = \phi^2_0/\phi^2_{\rm v} < 1$ together with
Eq.(\ref{eq:21}), caused by boundary conditions, the account for a
non-vanishing matter density between two parallel plates restricts
from above $0 \le \beta < \beta_{\rm max}$ the values of the
chameleon-matter coupling constant $\beta$. The coupling constant
$\beta_{\rm max}$ depends on a matter density and a distance between
plates and characterizes a maximal chameleon--matter coupling
constant, which can be measured for given experimental conditions.

\section{Acknowledgements}

We thank Clare Burrage for calling our attention to the paper
\cite{Burrage2016}. This work was supported by the Austrian ``Fonds
zur F\"orderung der Wissenschaftlichen Forschung'' (FWF) under 
contracts I689-N16, I862-N20 and P26781-N20.


\begin{thebibliography}{9}
\bibitem{Chameleon1} J. Khoury and A. Weltman, Phys. Rev. Lett. {\bf
  93}, 171104 (2004); Phys. Rev. D {\bf 69}, 044026 (2004).
\bibitem{Chameleon2}
D. F. Mota and D. J. Shaw, Phys. Rev. D {\bf 75}, 063501 (2007);
  Phys. Rev. Lett. {\bf 97}, 151102 (2007).
\bibitem{Will1993}
Cl. M. Will,
in {\it Theory and Experiment in Gravitational Physics}, 
Cambridge University Press, Cambridge 1993.
\bibitem{Zlatev1999}
I. Zlatev, L. Wang, and P. J. Steinhardt,
Phys. Rev. Lett. {\bf 82}, 896 (1999); 
P. J. Steinhardt, L. Wang, and I. Zlatev,
Phys. Rev. D {\bf 59}, 123504 (1999).
\bibitem{Tsujikawa2013}
S. Tsujikawa,
Class. Quantum Grav. {\bf 30}, 214003 (2013).
\bibitem{Ivanov2016a}
A. N. Ivanov and M. Wellenzohn,
{\it Can Chameleon Field be identified with Quintessence ?}, 
arXiv: 1607.00884 [gr-qc].
\bibitem{Perlmutter1997}
S. Perlmutter {\it et al.},
Bull. Am. Astron. Soc. {\bf 29}, 1351 (1997).
\bibitem{Riess1998}
A.  G. Riess {\it et al.},
Astron. J. {\bf 116}, 1009 (1998).
\bibitem{Perlmutter1999}
S. Perlmutter {\it et al.},
Astron. J. {\bf 517}, 565 (1999).
\bibitem{Goobar2000} A. Goobar {\it et al.}, Physica Scripta T {\bf
  85}, 47 (2000).  
\bibitem{Brax2004} Ph. Brax, C. van
  de  Bruck, A.--Ch. Devis, J. Khoury, and A. Weltman, Phys. Rev. D
  {\bf 70}, 123518 (2004).
\bibitem{Peebles2003}
P. J. E. Peebles and Bh. Ratra,
Rev. Mod. Phys. {\bf 75}, 559 (2003).
\bibitem{Copeland2006}
E. J. Copeland, M. Sami, S. Tsujikawa,
Int. J. Mod. Phys. D {\bf 15}, 1753 (2006).
\bibitem{Frieman2008}
J. A. Frieman, M. S. Turner, and Dr. Huterer,
Annu. Rev. Astron. Astrophys. {\bf 46}, 385 (2008).
\bibitem{Jain2013}
Bh. Jain {\it et al.}, arXiv:1309.5389 [astro-ph.CO].
\bibitem{Brax2015}
Ph. Brax and A.--C. Davis,
Phys. Rev. D {\bf 91}, 063503 (2015).
\bibitem{Pignol2015}
G. Pignol,
Int. J. Mod. Phys. A {\bf 30}, 1530048 (2015).
\bibitem{Abele2010}
H. Abele, T. Jenke, H. Leeb, and J. Schmiedmayer,
Phys.Rev. D {\bf 81}, 065019 (2010).
\bibitem{Jenke2011} 
T. Jenke, P. Geltenbort, H. Lemmel, and H. Abele,
  Nature Physics {\bf 7}, 468 (2011).
\bibitem{AbeleWF1}
H. Abele, T. Jenke, D. Stadler, and P. Geltenbort,
Nucl. Phys. A {\bf 827}, 593c (2009).
\bibitem{AbeleWF2}
T. Jenke, D. Stadler, H. Abele, and P. Geltenbort,
Nucl. Instr. and Meth. in Physics Res. A {\bf 611}, 318 (2009).
\bibitem{AbeleWF3}
H. Abele and H. Leeb,
New J. Phys. {\bf 14}, 055010 (2012).
\bibitem{Jenke2014} 
T. Jenke, G. Cronenberg, J. B\"urgdorfer,
  L. A. Chizhova, P. Geltenbort, A. N. Ivanov, T. Lauer, T. Lins,
  S. Rotter, H. Saul, U. Schmidt, and H. Abele, Phys. Rev. Lett. {\bf
    112}, 151105 (2014).
\bibitem{Brax2013}
Ph. Brax, G. Pignol, and D. Roulier,
Phys. Rev. D {\bf 88}, 083004 (2013).
\bibitem{Lemmel2015} 
H. Lemmel, Ph. Brax, A. N. Ivanov, T. Jenke,
  G. Pignol, M. Pitschmann, T. Potocar, M. Wellenzohn, 
M. Zawisky, and H. Abele, 
Phys. Lett. B {\bf 743}, 310 (2015).
\bibitem{Li2016}
K. Li, M. Arif, D. G. Cory, R. Haun,  B. Heacock, M. G. Huber,
J. Nsofini, D. A. Pushin, P. Saggu, D. Sarenac, C. B. Shahi, V. Skavysh,
W. M. Snow, and A. R. Young, Phys. Rev. D {\bf 93}, 062001 (2016).
\bibitem{Brax2011} 
Ph. Brax and G. Pignol, Phys. Rev. Lett. {\bf 107},
  111301 (2011).
\bibitem{Ivanov2013}
A. N. Ivanov, R. H\"ollwieser, T. Jenke, M. Wellenzohn, and H. Abele,
Phys. Rev. D {\bf 87}, 105013 (2013).
\bibitem{Burrage2014}
C. Burrage, E. J. Copeland, and  E. A. Hinds,
JCAP {\bf 03}, 042  (2015): arXiv:1408.1409v3 [astro-ph.CO].
\bibitem{Burrage2015} 
C. Burrage and E. J. Copeland, Contemporary
  Physics {\bf 57}, 164 (2015); arXiv:1507.07493 [astro-ph.CO].
\bibitem{Hamilton2015} 
P.  Hamilton, M. Jaffe, Ph. Haslinger,
  Q. Simmons, H. Müller, and J. Khoury,
Science {\bf 349}, 849 (2015).
\bibitem{Hamilton2016} B. Elder, J. Khoury, Ph. Haslinger, M. Jaffe,
  H. Müller, and P. Hamilton, {\it Chameleon Dark Energy and Atom
    Interferometry} e-Print: arXiv:1603.06587 [astro-ph.CO].
\bibitem{Gunther2016} G. Cronenberg, Invited talk at workshop on
  ``Dark energy in the  laboratory'' held on 20th - 22nd
  of April 2016 at Chicheley Hall,home of the Kavli Royal Society
International Centre, United Kingdom: http://nottingham.ac.uk/~ppzphy7/webpages/conferences/deitl
\bibitem{PDG2014} 
K. A. Olive {\it et al.} (Particle Data Group),
  Chin. Phys. A {\b 38}, 090001 (2014).
\bibitem{HMF1972}
{\it Handbook of Mathematical Functions With Formulas, Graphs, and
Mathematical Tables}, edited by M. Abramowitz and I. A. Stegun, 
Third Printing with corrections,
National Bureau of Standards, Applied Mathematics Series $\bullet$ 55,
p. 589 eq. 17.2.8, Washington  1972.
\bibitem{HMF1972a}
see Ref.~\cite{HMF1972} and p. 573 eq. 16.9.1 and p. 576 eq. 16.26.1
 and eq. 16.25.1.
\bibitem{HMF1972b}
see Ref.~\cite{HMF1972} p. 591 eq. 17.3.12.
\bibitem{Gibbs1975}
R. L. Gibbs, Am. J. Phys. {\bf 43}, 25 (1975).
\bibitem{Sedmik2016} 
R. Sedmik, Invited talk at workshop on
  ``Dark energy in the  laboratory'' held on 20th - 22nd
  of April 2016 at Chicheley Hall,home of the Kavli Royal Society
International Centre, United Kingdom; http://nottingham.ac.uk/~ppzphy7/webpages/conferences/deitl
\bibitem{Burrage2016}
Cl. Burrage, E. J. Copeland, and J. A. Stevenson,
arXiv:1604.00342 [astro-ph.CO].
\end{thebibliography}
\end{document}